%% file: main.tex
\documentclass[conference]{IEEEtran}
\usepackage[top=0.7in, bottom=0.96in, left=0.65in, right=0.65in]{geometry}
\input{input.tex}

\usepackage{epstopdf}
\usepackage{enumerate}
\usepackage{amsmath}
\usepackage{float}
\usepackage{color}
\usepackage{makeidx}
\usepackage{bm}
\usepackage{cleveref}
\usepackage{url}
\usepackage{steinmetz}
\usepackage{varwidth}
\usepackage{soul}
\usepackage{booktabs}
\usepackage{comment}
\usepackage{bbm}
\usepackage{subfigure}

\let\oldFootnote\footnote
\newcommand\nextToken\relax
\renewcommand\footnote[1]{%
	\oldFootnote{#1}\futurelet\nextToken\isFootnote}
\newcommand\isFootnote{%
	\ifx\footnote\nextToken\textsuperscript{,}\fi}

\DeclareMathOperator\RE{\mathrm{Re}}

\newcommand*{\E}{\mathrm{e}}

\def \rm {\mathrm}

\usepackage{algorithm}
\makeatletter
\newcommand{\algorithmname}{\ALG@name}
\renewcommand{\floatc@ruled}[2]{{\@fs@cfont #1:} #2\par}
\makeatother
\usepackage[noEnd=false]{algpseudocodex}
\tikzset{algpxIndentLine/.style={draw=black}}
\algrenewcommand{\alglinenumber}[1]{\bfseries\footnotesize #1}
\algrenewcommand{\textproc}{}
\algrenewcommand{\algorithmicrequire}{\textbf{Input:}}  
\algrenewcommand{\algorithmicensure}{\textbf{Output:}}

\begin{document}
    \title{ISAC with Backscattering RFID Tags: \\ Joint Beamforming Design}

    \author{Hao~Luo,~Umut Demirhan,~and~Ahmed~Alkhateeb\\School of Electrical, Computer, and Energy Engineering, Arizona State University\\Email: \{h.luo, udemirhan, alkhateeb\}@asu.edu}

    \maketitle

    \begin{abstract}	
        In this paper, we explore an integrated sensing and communication (ISAC) system with backscattering RFID tags. In this setup, an access point employs a communication beam to serve a user while leveraging a sensing beam to detect an RFID tag. Under the total transmit power constraint of the system, our objective is to design sensing and communication beams by considering the tag detection and communication requirements. First, we adopt zero-forcing to design the beamforming vectors, followed by solving a convex optimization problem to determine the power allocation between sensing and communication. Then, we study a joint beamforming design problem with the goal of minimizing the total transmit power while satisfying the tag detection and communication requirements. To resolve this, we re-formulate the non-convex constraints into convex second-order cone constraints. The simulation results demonstrate that, under different communication SINR requirements, joint beamforming optimization outperforms the zero-forcing-based method in terms of achievable detection distance, offering a promising approach for the ISAC-backscattering systems.
    \end{abstract}

    \section{Introduction} \label{sec:intro} 
    Integrated sensing and communication (ISAC) \cite{Demirhan2023a} has been envisioned to benefit various emerging applications in the future, with enhanced spectral, energy, and hardware efficiency. One potential scenario is the use of radio frequency identification (RFID) for inventory management in warehouses or retail stores, where low-cost passive RFID tags replace conventional barcodes. With ISAC, the systems can leverage sensing signals to track goods by detecting RFID tags, while simultaneously transmitting signals to the communication targets, e.g., wireless surveillance cameras or mobile devices, in a cooperative manner. Realizing such systems, however, is challenged by the limited detection range of RFID tags due to the absence of a built-in power source. This limitation could be mitigated, for instance, by employing MIMO beamforming at the RFID reader~\cite{Chen2016}. Moreover, the mutual interference between the sensing and communication signals may significantly impact the reading reliability of the RFID tags. Therefore, in this work, we investigate the joint sensing and communication beamforming design for the ISAC system with backscattering RFID tags. Recently, a relevant study~\cite{Galappaththige2023} introduced an integrated sensing and backscatter communication system, in which a basestation employs transmit beamforming to communicate with a user while broadcasting sensing signals to detect an RFID tag. Under different power allocation between sensing and communication, the authors analyzed the communication performance of both user and backscattering tag, and the sensing performance at the basestation. This work, however, neglected the consideration of the detection requirements of RFID tags, i.e., the sensitivities of both the tag and the reader, which is an important concern in practice. Furthermore, the study did not apply transmit beamforming to sensing signals, which is critical to extend the RFID detection range.

    In this paper, we investigate the joint beamforming design problem in an ISAC system with backscattering RFID tags. The contributions of this paper are summarized as follows:
    \begin{itemize}
        \item We formulate the ISAC-backscattering beamforming optimization problem to meet the requirements of tag detection and communication signal-to-inference-plus-noise ratio (SINR) under a total transmit power constraint.
        \item We, first, develop a zero-forcing based method. Specifically, we leverage zero-forcing to design the beamforming vectors, and solve a convex optimization problem to determine the transmit power allocation between sensing and communication beams.
        \item We, then, propose a joint design of sensing and communication beams while minimizing the transmit power. To address the non-convex constraints in the optimization problem, we transform them into second-order cone constraints. The re-formulated problem can be solved with the convex optimization tools.
    \end{itemize}
    The simulation results reveal that joint beamforming optimization can outperform the zero-forcing based approach in terms of the achievable detection distance, under different user SINR requirements. Also, thanks to the gain of beamforming, the achievable detection distance of the tag can be improved as the number of antennas increases.

    \section{System Model} \label{sec:sys}
    \begin{figure}[!t]
        \centering	
        \includegraphics[width=0.85\columnwidth]{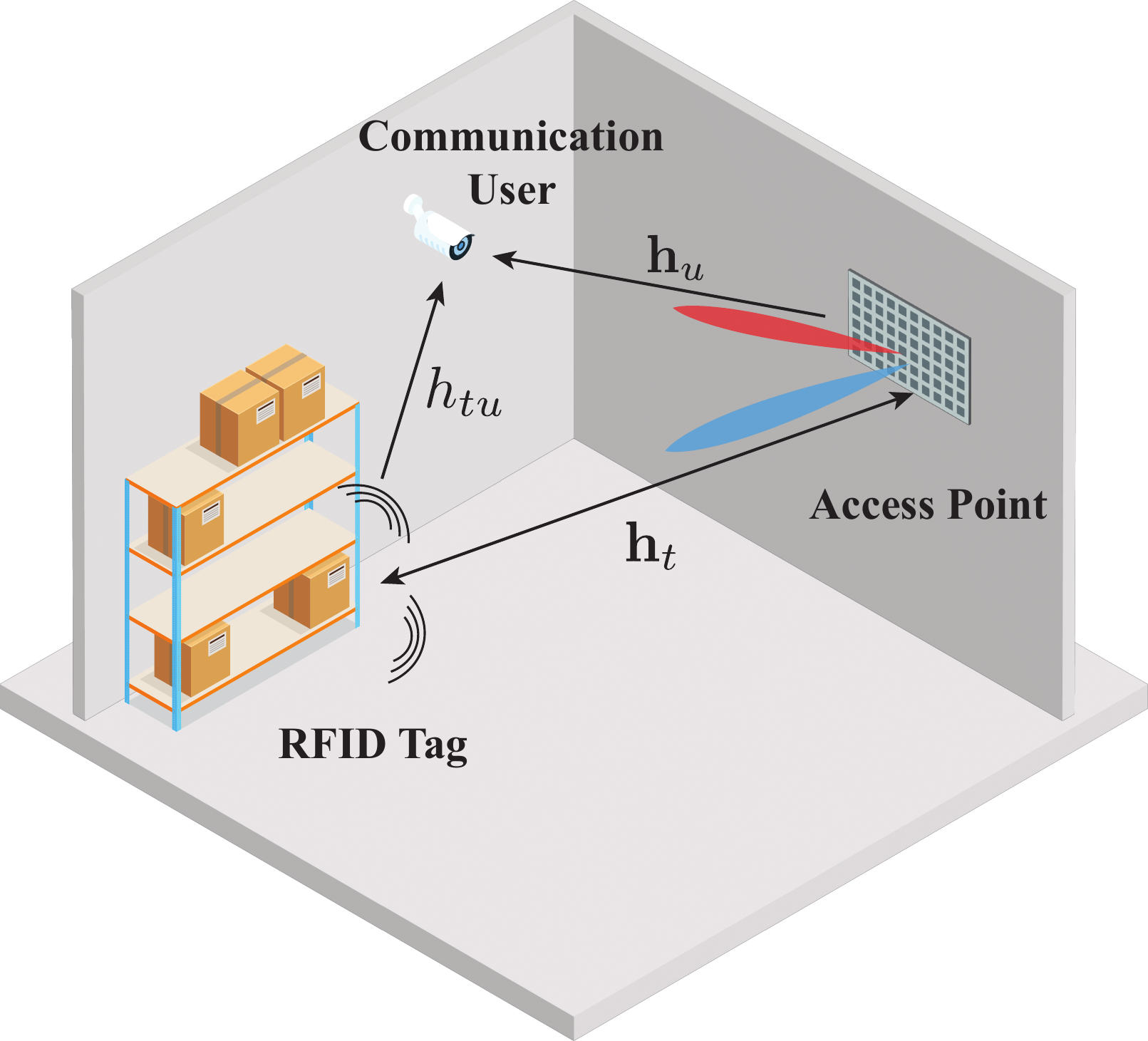}
        \caption{This figure illustrates the considered MIMO ISAC system, where an access point transmits sensing and communication waveforms to communicate with a user while detecting a passive RFID tag. This system could be applicable, for instance, in the context of RFID-aided inventory management and surveillance systems within a warehouse, as depicted in the figure.}	
        \label{fig:system_model}
    \end{figure}

    We consider a MIMO ISAC system with an access point, a passive RFID tag, and a communication user, as depicted in \figref{fig:system_model}. In this system, the access point transmits a joint communication and RFID sensing waveform to serve the user while simultaneously reading the tag. The tag scatters back the stored information by modulating the incident sensing signal. The backscattered signal received by the access point can be further processed for sensing purposes, such as positioning~\cite{Xu2023}. The access point is equipped with $N_t$ transmitter and $N_r$ receiver antennas. The tag and the user are each equipped with a single antenna. It is assumed that the access point can separate the received signal from the transmitted signal, with no consideration for signal leakage~\cite{Saad2014}.

    \subsection{Signal Model}
    The transmit signal at the access point is defined as the sum of the communication and sensing signals with corresponding beamforming. We denote the transmit signal as $\bx \in \bbC^{N_t}$, which can be formulated as follows
    \begin{equation}
        \bx = \bff_t s_t + \bff_u s_u,
    \end{equation} 
    where $s_t \in \bbC$ is the RFID sensing signal, e.g., continuous wave~\cite{EPCGlobal}, and $s_u \in \bbC$ is the data for the communication user. The signals are assumed to be unit average energy, i.e., $\bbE[|s_t|^2]=\bbE[|s_u|^2]=1$. $\bff_t\in\bbC^{N_t}$ and $\bff_u \in \bbC^{N_t}$ are sensing and communication beamforming vectors, respectively. The beamforming vectors satisfy the transmit power constraint, i.e., $\|\bff_t\|^2 + \|\bff_u\|^2 \leq P$, where $P$ is the total transmit power.

    \subsection{Backscatter Model}
    The tag leverages the received signals to perform backscatter modulation. We denote the channel between the access point and the tag as $\bh_t \in \bbC^{N_t}$. The received signal at the tag can be written as 
    \begin{align} \label{eq:recv_sig_tag}
        y_t &= \bh_t^H \bx + n_t \nonumber \\
        &= \bh_t^H \bff_t s_t + \bh_t^H \bff_u s_u + n_t,
    \end{align}
    where $n_t \sim \cC\cN(0, \sigma_t^2)$ is the receiver noise at the tag. Then, the tag scatters back the signal that contains the stored data by modulating the impinging signal. Formally, the backscatter-modulated signal can be expressed as 
    \begin{equation}
        r = \sqrt{\eta} \, y_t \, d,
    \end{equation}
    where $\eta$ is the backscatter modulation efficiency, and $d$ is the encoded tag's data with $\bbE[|d|^2]=1$. Next, the signal modulated by the tag is received at the access point. We assume that the channel between the access point and the tag is reciprocal. Then, the received signal at the access point is given as 
    \begin{align} \label{eq:recv_sig_reader}
        y_r &= \bw^H \bh_t \, r + \bw^H \bn_r \nonumber \\
        &= \bw^H \bh_t \sqrt{\eta} \, d (\bh_t^H \bff_t s_t + \bh_t^H \bff_u s_u + n_t) + \bw^H \bn_r,
    \end{align}
    where $\bw \in \bbC^{N_r}$ is the combining vector used by the access point, and $\bn_r \sim \cC\cN(0, \sigma_r^2 \bI)$ is the receiver noise vector. Given the received signal at the tag \eqref{eq:recv_sig_tag}, the corresponding received SINR can be obtained as
    \begin{align}
        \rm{SINR}_t &= \frac{\bbE\left[|\bh_t^H \bff_t s_t|^2\right]}{\bbE\left[|\bh_t^H \bff_u s_u|^2\right] + \bbE\left[|n_t|^2\right]} \nonumber \\
        &= \frac{|\bh_t^H \bff_t|^2}{|\bh_t^H \bff_u|^2 + \sigma_t^2} \nonumber \\
        &= \frac{P_t|\bh_t^H \bar{\bff}_t|^2}{P_u|\bh_t^H \bar{\bff}_u|^2 + \sigma_t^2},
    \end{align}
    where $P_t$ and $P_u$ are the allocated power for sensing and communication. $\bar{\bff}_t$ and $\bar{\bff}_u$ are the normalized sensing and communication beamforming vectors, i.e., $\bar{\bff}_t=\bff_t/\|\bff_t\|$, $\bar{\bff}_u=\bff_u/\|\bff_u\|$. Similarly, based on \eqref{eq:recv_sig_reader}, the received SINR at the access point is given by
    \begin{align}
        \rm{SINR}_r &= \frac{\eta |\bw^H \bh_t|^2 |\bh_t^H \bff_t|^2}{\eta |\bw^H \bh_t|^2 |\bh_t^H \bff_u|^2 + \eta \, \sigma_t^2 |\bw^H \bh_t|^2 + \sigma_r^2} \nonumber \\
        &= \frac{\eta |\bw^H \bh_t|^2 P_t|\bh_t^H \bar{\bff}_t|^2}{\eta |\bw^H \bh_t|^2 P_u|\bh_t^H \bar{\bff}_u|^2 + \eta \, \sigma_t^2 |\bw^H \bh_t|^2 + \sigma_r^2}.
    \end{align}
    Note that, to detect the tag successfully, it is essential for both SINRs at the tag and the access point to meet their respective sensitivity constraints.

    \subsection{Communication Model}
    For the communication, the user receives the transmitted signal from the access point and the backscattered signal from the tag. Thus, the received signal at the user is given by
    \begin{align}
        y_u &= \bh_u^H \bx + h_{tu} r + n_u \nonumber \\
        &= \bh_u^H \bff_u s_u + \bh_u^H \bff_t s_t \nonumber \\
        & \quad + h_{tu} \sqrt{\eta} \, d(\bh_t^H \bff_t s_t + \bh_t^H \bff_u s_u + n_t) + n_u,
    \end{align}
    where $\bh_u \in \bbC^{N_t}$ is the channel between the access point and the user, $h_{tu} \in \bbC$ is the channel between the tag and the user, and $n_u \sim \cC\cN(0, \sigma_u^2)$ is the receiver noise at the user. Then, the received SINR at the user is given in \eqref{eq:SINR_user}.
    \begin{figure*}[!t]
        \begin{align} \label{eq:SINR_user}
            \rm{SINR}_u &= \frac{|\bh_u^H \bff_u|^2}{|\bh_u^H \bff_t|^2 + \eta |h_{tu}|^2 (|\bh_t^H \bff_t|^2 + |\bh_t^H \bff_u|^2 + \sigma_t^2) + \sigma_u^2} \nonumber \\
            &= \frac{P_u|\bh_u^H \bar{\bff}_u|^2}{P_t|\bh_u^H \bar{\bff}_t|^2 + \eta |h_{tu}|^2 (P_t|\bh_t^H \bar{\bff}_t|^2 + P_u|\bh_t^H \bar{\bff}_u|^2 + \sigma_t^2) + \sigma_u^2}.
        \end{align}
    \end{figure*}

    \section{Problem Formulation} \label{sec:prob}
    In this work, our objective is to design a sensing beam $\bff_t$ and a communication beam $\bff_u$ that achieve tag detection while meeting the SINR requirement of the user. It is worth noting that, in this work, we focus on the beamforming design problem under the assumption that the communication channel and the tag's position are known to the access point. Also, the combining vector of the access point is assumed to be equal-gain combining, i.e., $\bw = \bh_t/\|\bh_t\|$. Then, the problem can be formulated as the following feasibility problem
    \begin{subequations} \label{eq:opt_feas}
        \begin{align}
            \rm{find} \quad & \{\bff_t, \bff_u\} \\
            \textrm{s.t.} \quad & \rm{SINR}_u \geq \gamma_u, \label{eq:opt_feas_constraint_1} \\
            \quad & \rm{SINR}_t \geq \gamma_t, \label{eq:opt_feas_constraint_2} \\
            \quad & \rm{SINR}_r \geq \gamma_r, \label{eq:opt_feas_constraint_3} \\
            \quad & \|\bff_t\|^2 + \|\bff_u\|^2 \leq P.
        \end{align}
    \end{subequations}
    In this problem, \eqref{eq:opt_feas_constraint_1} guarantees that the user SINR requirement $\gamma_u$ is satisfied. \eqref{eq:opt_feas_constraint_2} and \eqref{eq:opt_feas_constraint_3} aim to ensure that the tag will be read by the access point, taking into account both the tag's sensitivity $\gamma_t$ and the reader's sensitivity $\gamma_r$. In the next section, we introduce the proposed beamforming schemes.

    \section{Proposed Solutions} \label{sec:sol}
    In this section, we first present the zero-forcing based beamforming with power allocation optimization. Then, we introduce the joint design of sensing and communication beams.

    \subsection{Zero-Forcing Beamforming with Power Allocation Optimization}
    In this subsection, we propose to leverage zero-forcing to design the beamforming vectors, and then optimize the transmit power allocation between sensing and communication. For the zero-forcing solution~\cite{Bjornson2014}, in this work, we project the tag's channel on the null-space of the user's channel, and vice versa. Let $\bH = [\bh_t, \bh_u]$ contain the channels of the tag and the user. The zero-forcing beamforming vectors are given by
    \begin{equation}
        [\bff_t^\rm{ZF}, \bff_u^\rm{ZF}] = \bH (\bH^H \bH)^{-1}.
    \end{equation}
    The beamforming vectors can then be normalized to satisfy the total power constraint, i.e., $\bff_t = \sqrt{P_t} \, (\bff_t^\rm{ZF} / \|\bff_t^\rm{ZF}\|), \bff_u = \sqrt{P_u} \, (\bff_u^\rm{ZF} / \|\bff_u^\rm{ZF}\|)$. The feasibility problem in \eqref{eq:opt_feas} may have multiple solutions, and the most desirable is the one with the minimal power. For this purpose, we can replace the objective of the feasibility problem with the power minimization, and formulate the following problem
    \begin{subequations} \label{eq:opt_PA}
        \begin{align}
            \min_{P_t, P_u} \quad & P_t + P_u \\
            \textrm{s.t.} \quad & \rm{SINR}_u \geq \gamma_u, \\
            \quad & \rm{SINR}_t \geq \gamma_t, \\
            \quad & \rm{SINR}_r \geq \gamma_r, \\
            \quad & P_t + P_u \leq P.
        \end{align}
    \end{subequations}
    Since the beamforming vectors have been determined by the zero-forcing, this problem is a linear programming and can be solved by the convex solvers, e.g., CVX~\cite{cvx}.

    \subsection{Joint Beamforming Optimization}
    In this subsection, we explore a more desirable approach for the same problem. Specifically, we aim to jointly optimize the beamforming and the power allocation for sensing and communication. To start, we formulate the joint beamforming optimization problem as follows
    \begin{subequations} \label{eq:opt_BF}
        \begin{align}
            \min_{\bff_t, \bff_u} \quad & \|\bff_t\| + \|\bff_u\| \\
            \textrm{s.t.} \quad & \rm{SINR}_u \geq \gamma_u, \label{eq:opt_BF_constraint_1} \\
            \quad & \rm{SINR}_t \geq \gamma_t, \label{eq:opt_BF_constraint_2} \\
            \quad & \rm{SINR}_r \geq \gamma_r, \label{eq:opt_BF_constraint_3} \\
            \quad & \|\bff_t\|^2 + \|\bff_u\|^2 \leq P.
        \end{align}
    \end{subequations}
    Since the optimization variables in the SINR constraints are in fractional forms, the problem in \eqref{eq:opt_BF} is non-convex. Notably, the SINR constraints can be transformed into second-order cone constraints~\cite{Demirhan2023b}, which are convex. For instance, we can first re-write the user's SINR constraint in \eqref{eq:opt_BF_constraint_1} as follows
    \begin{align} \label{eq:opt_BF_constraint_1_rewritten}
        &\frac{1}{\gamma_u} |\bh_u^H \bff_u|^2 \nonumber \\
        &\geq |\bh_u^H \bff_t|^2 + \eta |h_{tu}|^2 (|\bh_t^H \bff_t|^2 + |\bh_t^H \bff_u|^2 + \sigma_t^2) + \sigma_u^2.
    \end{align}
    The absolute on the left-hand side, however, is a non-linear function. To address this, we utilize the observation that arbitrary phase rotation can be added to the expression in an absolute without affecting the value, i.e., $|\bh_u^H \bff_u| = |\bh_u^H \bff_u \E^{j \theta}|$. Without loss of optimality, we can select a phase rotation $\E^{j \theta}$ such that $\bh_u^H \bff_u$ becomes real and positive, i.e.,
    \begin{align}
        |\bh_u^H \bff_u| = \RE\left\{\bh_u^H \bff_u\right\}.
    \end{align}
    Then, by taking the square root of both sides in \eqref{eq:opt_BF_constraint_1_rewritten}, the SINR constraint of the user can be cast in the second-order cone form, given by
    \begin{equation} \label{eq:reformulated_SINR_u}
        \sqrt{\frac{1}{\gamma_u}} \textrm{Re}\left\{ \bh_u^H \bff_u \right\} \geq
        \begin{Vmatrix}
            \bh_u^H \bff_t\\
            \sqrt{\eta} \, h_{tu} \, \bh_t^H \bff_t \\
            \sqrt{\eta} \, h_{tu} \, \bh_t^H \bff_u \\
            \sqrt{\eta} \, h_{tu} \, \sigma_t \\
            \sigma_{u}
        \end{Vmatrix}.
    \end{equation}
    In a similar way, the sensitivity constraints in \eqref{eq:opt_BF_constraint_2} and \eqref{eq:opt_BF_constraint_3} can be re-written as follows
    \begin{equation} \label{eq:reformulated_SINR_t}
        \sqrt{\frac{1}{\gamma_t}} \textrm{Re}\left\{ \bh_t^H \bff_t \right\} \geq
        \begin{Vmatrix}
            \bh_t^H \bff_u\\
            \sigma_{t}
        \end{Vmatrix},
    \end{equation}

    \begin{equation} \label{eq:reformulated_SINR_r}
        \sqrt{\frac{\eta}{\gamma_r}} |\bw^H \bh_t| \textrm{Re}\left\{ \bh_t^H \bff_t \right\} \geq
        \begin{Vmatrix}
            \sqrt{\eta} \bw^H \bh_t \bh_t^H \bff_u\\
            \sqrt{\eta} \bw^H \bh_t \sigma_t \\
            \sigma_{r}
        \end{Vmatrix}.
    \end{equation}
    With the reformulation, the SINR constraints in \eqref{eq:reformulated_SINR_u}-\eqref{eq:reformulated_SINR_r} are equivalent to \eqref{eq:opt_BF_constraint_1}-\eqref{eq:opt_BF_constraint_3}.
    Therefore, the problem in \eqref{eq:opt_BF} becomes a convex second-order cone programming, which can be solved with the convex solvers.

    \section{Simulation Results} \label{sec:sim}
    In this section, we first describe the simulation setup and then evaluate the performance of zero-forcing based method and joint beamforming optimization.

    \subsection{Simulation Setup}
    We consider a scenario where the access point is placed at the origin in Cartesian coordinates. The transmit and receive antennas of the access point are uniform linear arrays along y-axis, looking at the positive direction of x-axis. The operating frequency is $2.4$ GHz, and the spacing between antennas is half wavelength. The total transmit power of the access point is set to $P=30$ dBm. For the channels, we adopt a line-of-sight channel model, and the variance of receive noise is $\sigma_t^2=\sigma_r^2=\sigma_u^2=10\log_{10}(kTB)+N_f$ dBm, where $k$ is Boltzmann's constant, $T=270$ Kelvin, $B=10$ MHz, and $N_f=7$ dB is the noise figure. According to the datasheets~\cite{Impinj_Tag,Impinj_Reader}, the tag's and reader's sensitivity values are set to $-25.5$ dBm and $-94$ dBm, respectively. The backscatter-modulation efficiency of the tag is set to $\eta = 0.16$ by assuming a given differential radar cross section~\cite{Impinj_Tag} and FM0 encoding scheme~\cite{Nikitin2008}.

    \subsection{Performance Evaluation}
    \begin{figure}[!t]
        \centering
        
        \subfigure[$\rm{SINR}_u=0$ dB]{
            \includegraphics[width=0.77\columnwidth]{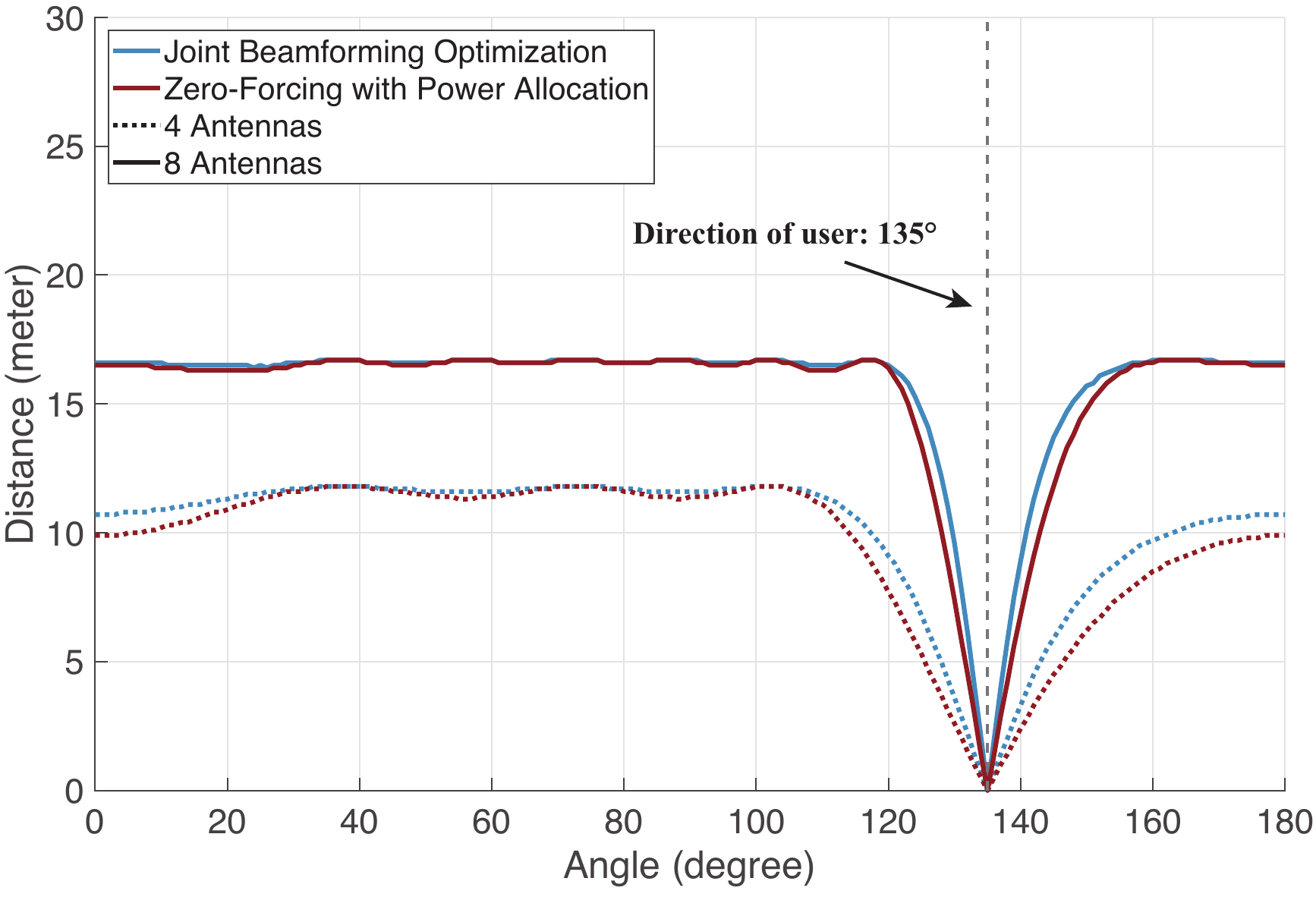}
            \label{fig:max_distance_low_SINR}
        }
        \hfill 
        \subfigure[$\rm{SINR}_u=10$ dB]{
            \includegraphics[width=0.77\columnwidth]{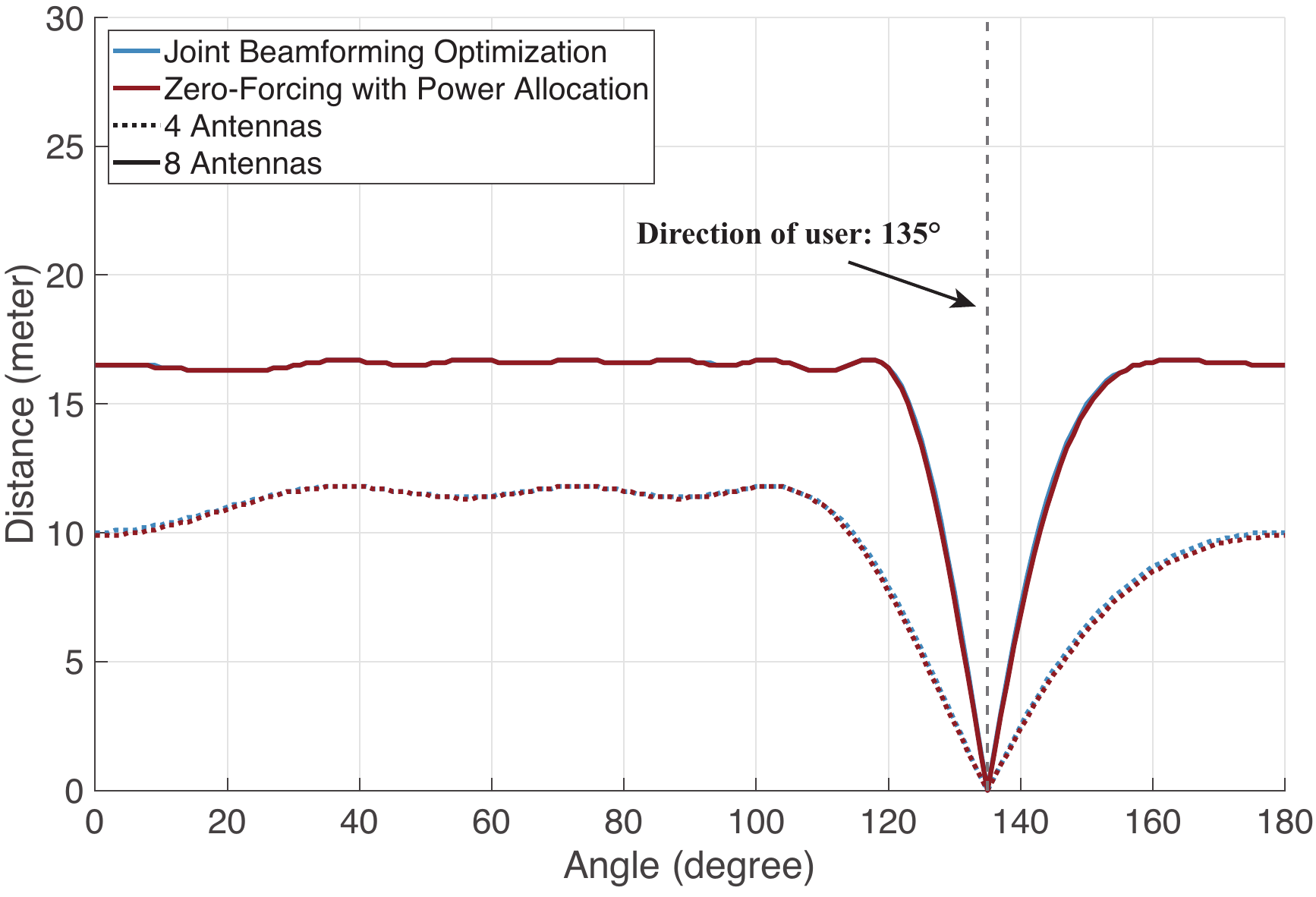}
            \label{fig:max_distance_high_SINR}
        }
        
        \caption{The achievable detection distance of different tag directions, where the user's position remains fixed at $(5/\sqrt{2}, 5/\sqrt{2})$. (a) and (b) show the performance under low and high user SINR requirements, respectively.}	
        \label{fig:max_distance}
    \end{figure}

    \begin{figure}[!t]
        \centering	
        \includegraphics[width=0.77\columnwidth]{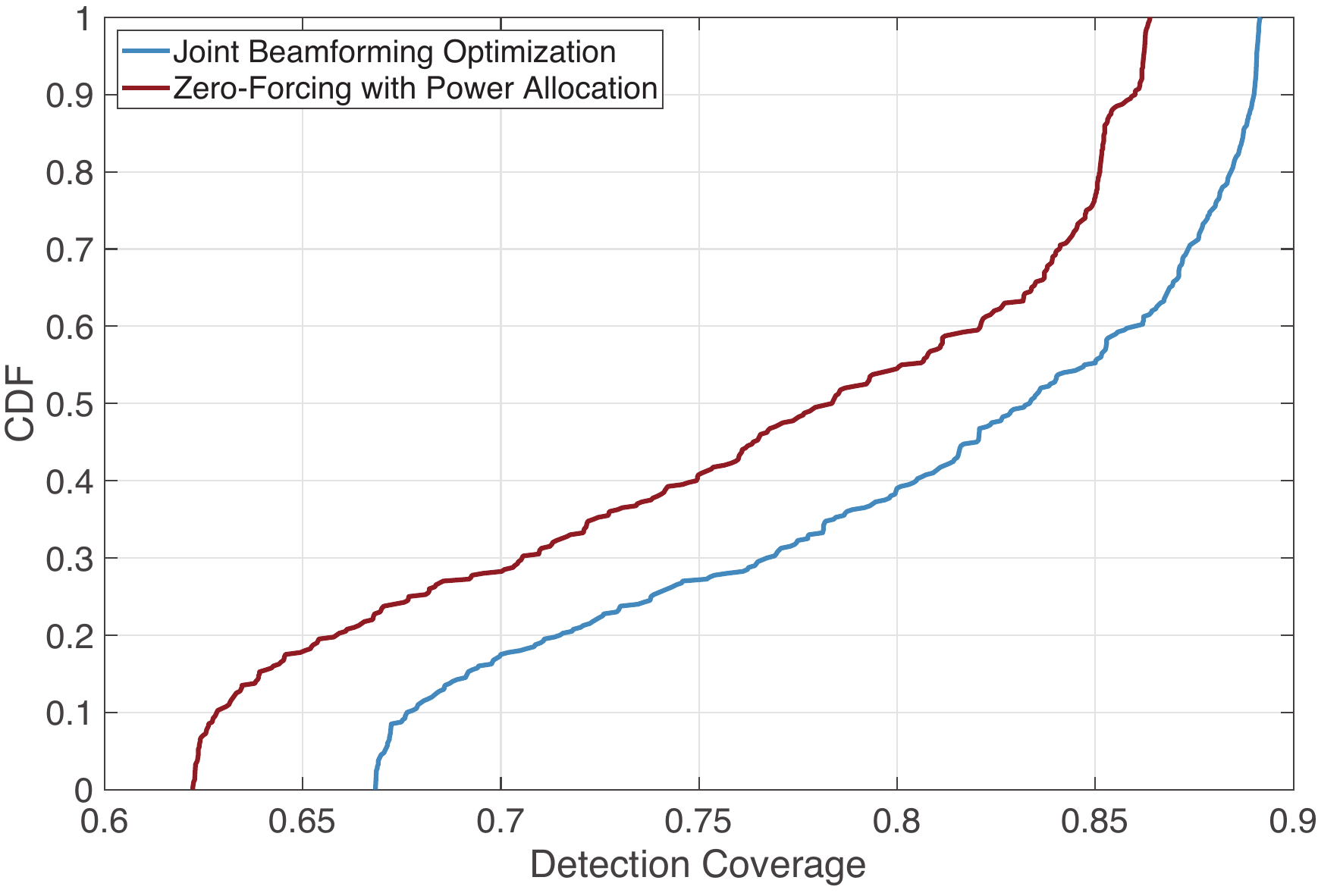}
        \caption{The CDF of the coverage ratio across different user positions. At each user position, the detection coverage is computed as the average ratio of the achievable detection distance to the upper bound across different angles. In this simulation, the number of antennas is $4$, and $\rm{SINR}_u$ is set to $0$ dB.}	
        \label{fig:max_distance_CDF}
    \end{figure}

    In \figref{fig:max_distance_low_SINR} and \figref{fig:max_distance_high_SINR}, we evaluate the achievable detection distance of different tag directions under two different user SINR requirements, i.e., low SINR and high SINR. The user is positioned at coordinates $(5/\sqrt{2}, 5/\sqrt{2})$, i.e., at 135 degrees angle to the access point and 5 meters away. As it can been seen in the figures, the achievable detection distance increases with the growing number of antennas, thanks to the gain provided by beamforming. When the tag and user directions are close, however, the achievable detection distance decreases due to high interference. In \figref{fig:max_distance_low_SINR}, under a low user SINR requirement, the joint beamforming optimization outperforms the zero-forcing based method. This is because the user can tolerate higher interference, making the zero-forcing method less optimal. As for the high user SINR requirement, the zero-forcing based method and the joint beamforming optimization demonstrate comparable performance, as depicted in \figref{fig:max_distance_high_SINR}. Furthermore, in \figref{fig:max_distance_CDF}, we present the cumulative distribution function (CDF) of the detection coverage of tag detection among different user positions. In this simulation, we uniformly sample $400$ user positions, with x coordinates  drawn from the range $[0, 20]$ and y coordinates drawn from the range $[-20, 20]$. At each user position, the detection coverage is computed as the average ratio of the achievable detection distance to the upper bound across different angles. The upper bound is defined as the achievable detection distance of the tag when there is no communication user. Under the scenario of low user SINR requirement, we can observe that joint beamforming optimization can provide better coverage compared to the zero-forcing based method. This highlights the gain of joint beamforming optimization.

\begin{figure}[!t]
        \centering	
        \includegraphics[width=0.77\columnwidth]{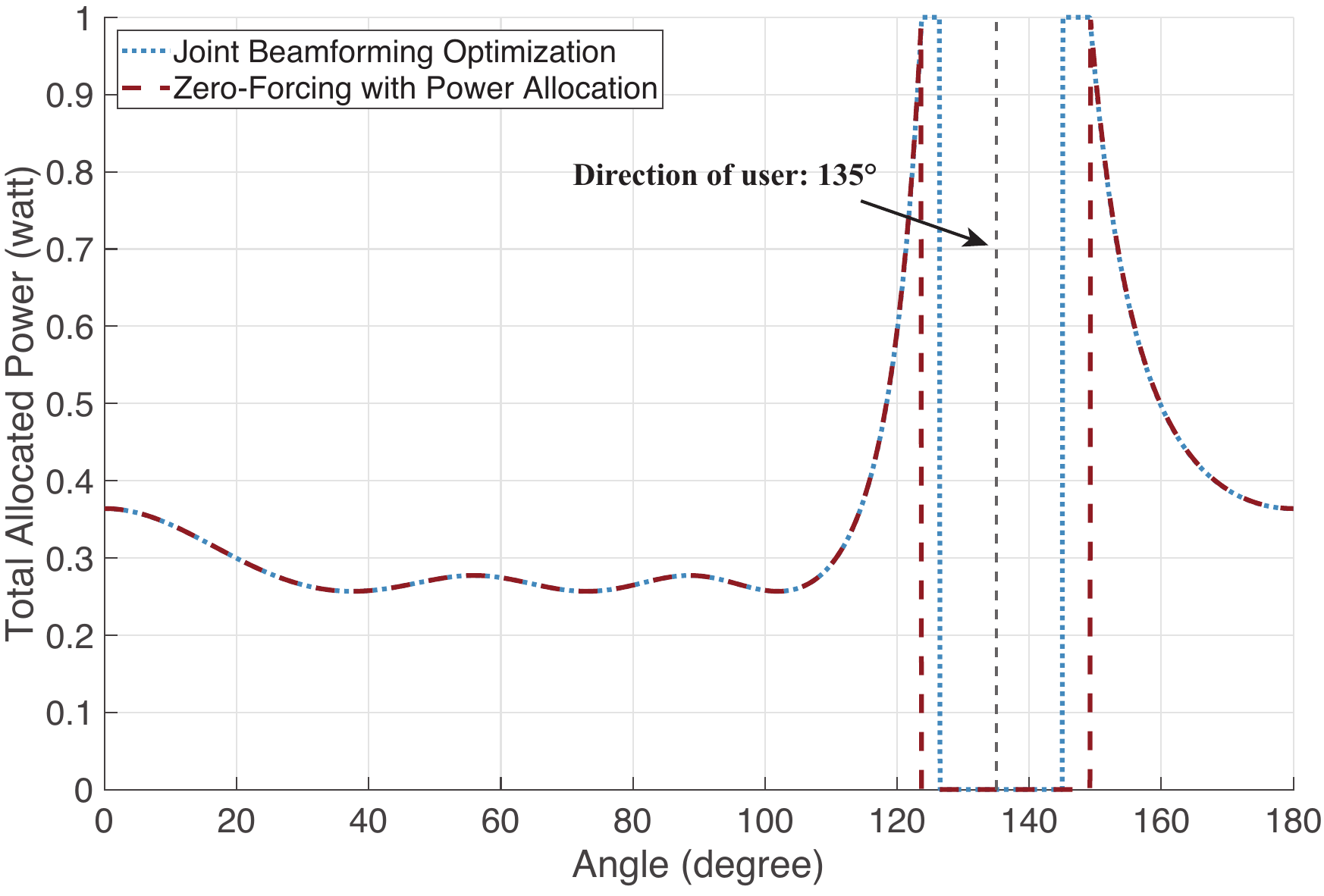}
        \caption{The total allocated power with different tag directions, where the user's position is at $(5/\sqrt{2}, 5/\sqrt{2})$, and the tag's distance is fixed at $6$ meters. In this simulation, the number of antennas is $4$, and $\rm{SINR}_u$ is set to $0$ dB.}	
        \label{fig:transmit_power}
    \end{figure}

\begin{figure}[!t]
        \centering	

        \subfigure[Joint beamforming optimization]{
            \includegraphics[width=0.77\columnwidth]{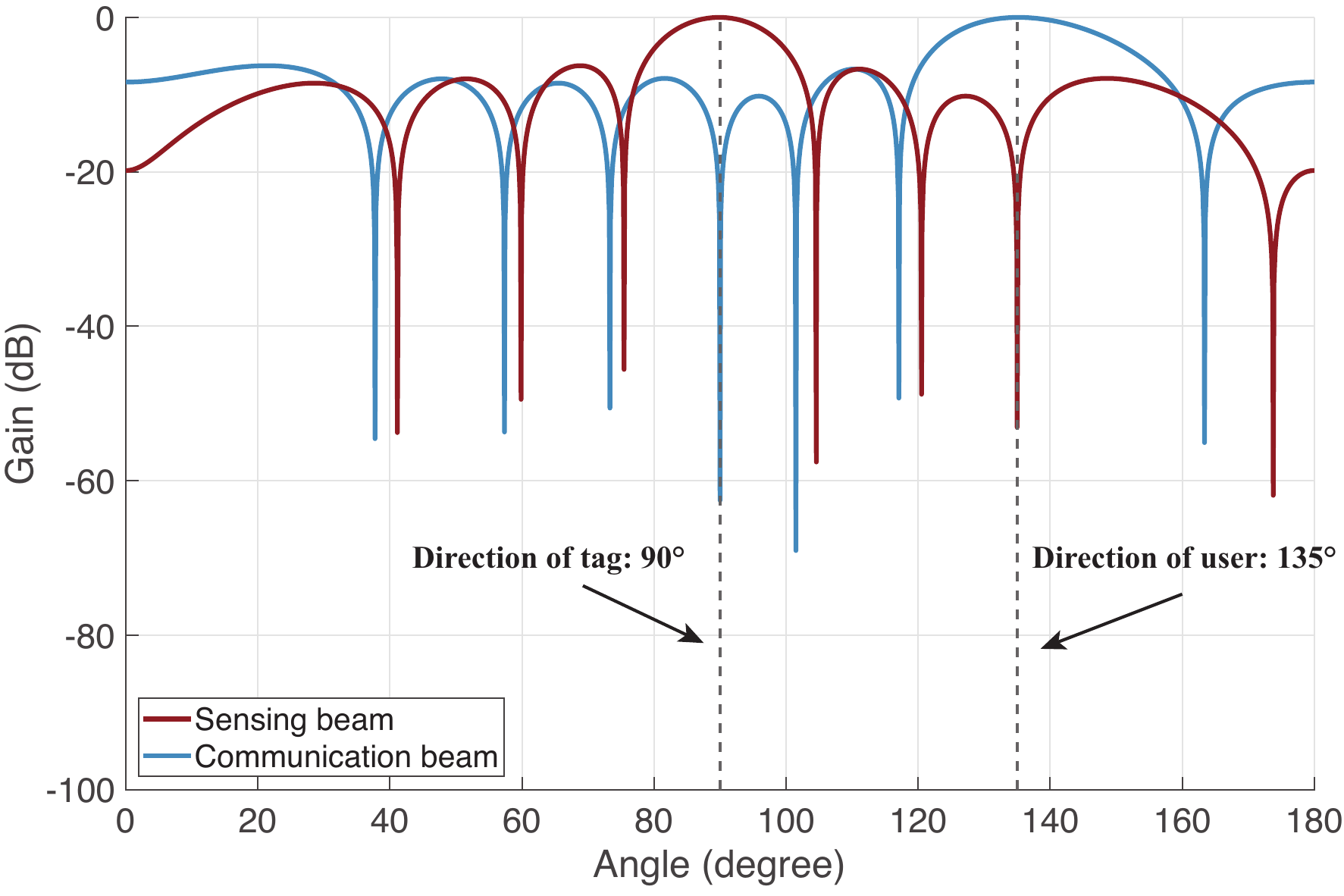}
            \label{fig:beam_pattern_BF}
        }
        \hfill 
        \subfigure[Zero-forcing with power allocation optimization]{
            \includegraphics[width=0.77\columnwidth]{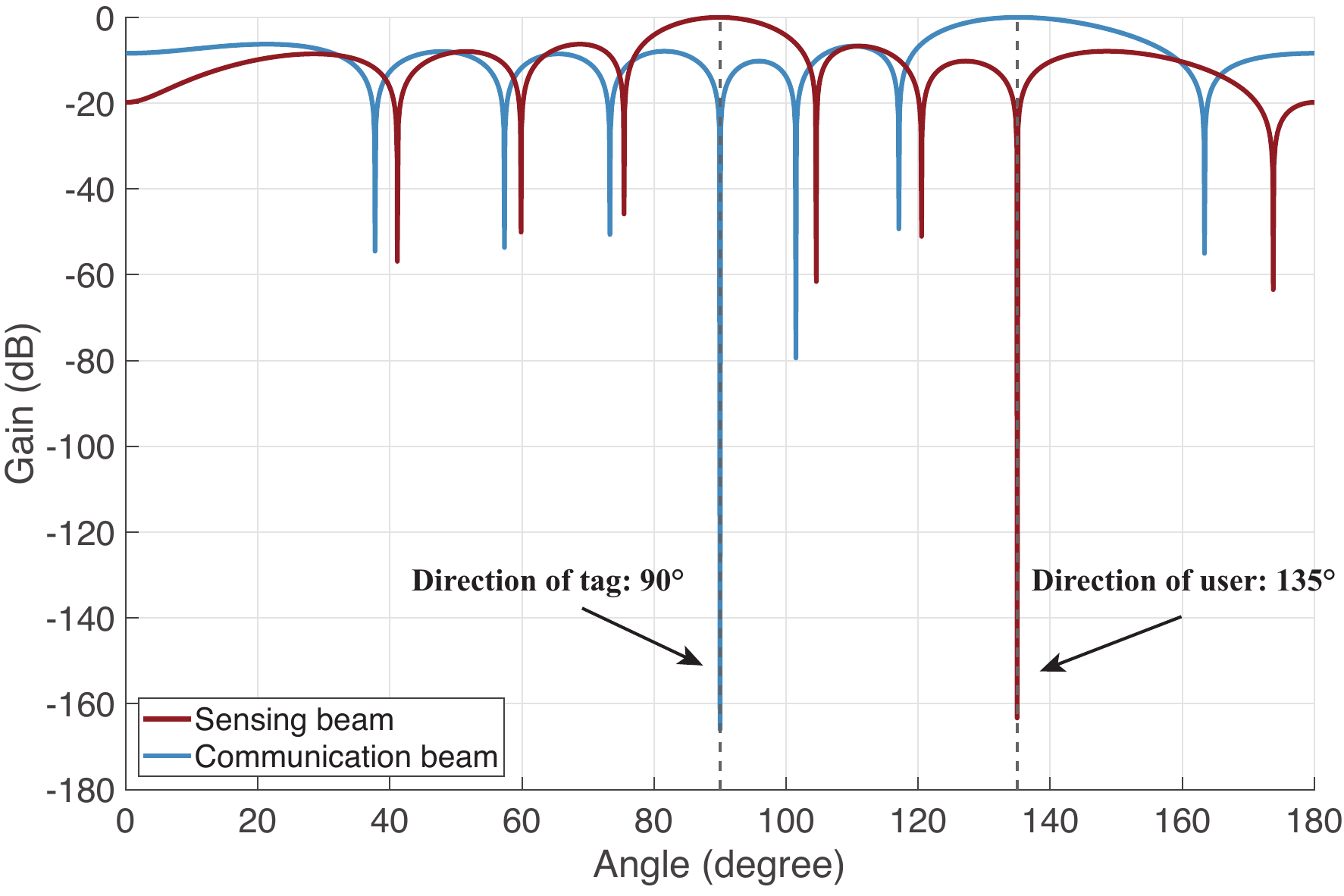}
            \label{fig:beam_pattern_ZF}
        }

        \caption{The beamforming patterns of the proposed solutions. The number of antennas is $8$. The directions of the tag and the user remain fixed at $90^{\circ}$ and $135^{\circ}$. (a) shows the beamforming pattern of joint beamforming optimization, and (b) depicts the beamforming pattern of zero-forcing based method.}	
        \label{fig:beam_pattern}
    \end{figure}

    In \figref{fig:transmit_power}, we evaluate the total allocated power of the proposed solutions with varying tag directions. In this simulation, the user is positioned at coordinates $(5/\sqrt{2}, 5/\sqrt{2})$, and the distance between the tag and the access point remain fixed at $6$ meters. In this figure, when the power is zero, it indicates that there is no feasible solution. As shown in the figure, when the tag is closer to the user, the total allocated power increases until it reaches the maximum limit of total transmit power. This is because more transmit power is required to meet the requirements of the user SINR and the tag detection, given higher interference. In \figref{fig:beam_pattern_BF}, we plot the beam patterns from the results of joint beamforming optimization. We can observe that the main lobes of sensing and communication beams are steered toward the directions of the tag and the user, respectively. Meanwhile, the sensing beam forms a null at the direction of the user in order to eliminate the interference. Similarly, a null is shaped at the direction of the tag by the communication beam. Compared to the zero-forcing based method, as depicted in \figref{fig:beam_pattern_ZF}, joint beamforming optimization does not require the reduction of interference to a very low level, which allows for more freedom in the beamforming design.

    \section{Conclusion} \label{sec:conclusion}
    In this work, we investigate the beamforming design for an ISAC system with backscattering RFID tags. Specifically, we formulate a beamforming design problem that satisfies the total transmit power constraint while meeting the tag detection and communication requirements. To solve this problem, we first propose to design beamforming vectors using zero-forcing, and then determine the power allocation on sensing and communication. Next, we develop a joint beamforming design with the aim of minimizing the total transmit power. Due to the non-convexity of the problem constraints, we re-formulate them into convex second-order constraints. The results highlights the efficacy of joint beamforming optimization in terms of achievable detection distance under different communication SINR requirements.

    \section*{Acknowledgement}
	This work is supported by the National Science Foundation under Grant No. 2229530.

\end{document}

%% file: input.tex
\usepackage{amsfonts}
\usepackage{times}
\usepackage{graphicx}
\usepackage{latexsym}
\usepackage{dsfont}
\usepackage{amssymb}
\usepackage{amsmath}
\usepackage{cite}
\usepackage{verbatim}

\newcommand{\figref}[1]{{Fig.}~\ref{#1}}

\def\bb0{{\mathbb{0}}}

\def\bb{{\mathbf{b}}}

\def\bff{{\mathbf{f}}}

\def\bh{{\mathbf{h}}}

\def\bn{{\mathbf{n}}}

\def\bw{{\mathbf{w}}}
\def\bx{{\mathbf{x}}}

\def\b0{{\mathbf{0}}}

\def\bH{{\mathbf{H}}}
\def\bI{{\mathbf{I}}}

\def\bbC{{\mathbb{C}}}

\def\bbE{{\mathbb{E}}}

\def\cC{\mathcal{C}}

\def\cN{\mathcal{N}}

\def\sf0{{\mathsf{0}}}

